\documentstyle[prb,aps,multicol]{revtex}
\renewcommand{\narrowtext}{\begin{multicols}{2} \global\columnwidth20.5pc}
\renewcommand{\widetext}{\end{multicols} \global\columnwidth42.5pc}

\begin{document}
\title{On the critical temperature of two-channel Kondo model: Reply
to  cond-mat/0009283}
\author{I.~L. Aleiner$^{1}$, B.~L. Altshuler$^{2}$, and
Y. M. Galperin$^{3}$}  
\address{$^{1}$ Department of Physics and Astronomy, SUNY at Stony Brook, 
Stony Brook, NY 11794, USA,\\ 
$^{3}$Physics Department, Princeton University, Princeton, NJ 08544,
USA and NEC Research Institute,\\ 
4 Independence Way, Princeton, NJ 08540, USA\\ 
$^3$Department of Physics, University of Oslo, P. O. Box
1048, N-0316 Oslo, Norway and
Division of Solid State Physics, \\ Ioffe
Institute of the Russian Academy of Sciences,
St. Petersburg 194021, Russia } 
\date{February 28, 2001}
\maketitle
\begin{abstract}

We discuss the comment by Zar{\'a}nd and Zawadowski (cond-mat/0009283)
on our preprint cond-mat/0007430 where it has been shown that the
strong coupling regime for the two-channel Kondo model  of two level
system can be never realized for any realistic microscopic
description. The authors of the comment state that the Kondo
temperature can be substantially increased due to electron-hole
asymmetry. Here we show by direct calculation that the electron-hole
asymmetry does not enter the leading logarithmic
approximation. Consequently, we disagree with the aforementioned
comment.

\end{abstract}

\narrowtext 
Hereby we discuss the comment by Zar{\'a}nd and
Zawadowski~\cite{zzcom} regarding our preprint~\cite{ourcom} on
the { Kondo} temperature, $T_K$, of the two-channel Kondo
model based on a heavy particle in a double-well potential (DWP)
(see Ref.~\onlinecite{CZ,2CK} for a review).  A serious
deficiency of the original 2 channel Kondo (2CK) model was too
small tunneling constant~\cite{zz1} calculated in leading
logarithmic approximation { turns out to be below $0.1$ K }
{ Authors of} Ref.~\onlinecite{zz1} { attempted to ``cure''
  the situation by including} virtual processes via the {
  third} excited state { in DWP in hope that it would}
increase the tunneling amplitude.

{ In Ref.~\onlinecite{ourcom} we demonstrated that
when {\em all of the exited states} are included
{\it the Kondo temperature gets reduced even further.}
It turns out that the Kondo} logarithmic singularity 
{ should be cut-off by} 
$\varepsilon_3 \sim \hbar t_{\text{tun}}\ll \varepsilon_F$
Here $\varepsilon_3$ is the 
{ spacing} between the second and the third levels { in DWP}, 
$t_{\text{tun}}$ it the tunneling time,
{ and} $\varepsilon_F$ denotes the Fermi energy. 
Note that long ago Kagan and Prokof'ev made a similar statement
in connection with diffusion of heavy particles in metals.  Since
$T_K$ is proportional to the cut-off and $\varepsilon_3 \ll
\varepsilon_F$, the Kondo temperature turns out to be hopelessly
small { for the strong coupling 2CK limit to be achieved
  experimentally}.  Moreover, we found that the spacing between
first two levels always exceeds $T_K$.  This prevents realization
of this strong coupling regime even in principle.~\cite{kp}
In Ref.~\onlinecite{ourcom} { we emphasized the importance of
  taking into account all of the exited states
as well as the fact} that the exponential smallness of the 
tunneling probability cannot { disappear}.

Formally, this smallness follows from the cancellation of virtual
contributions of different excited states.  At the same time, the
dramatic reduction of the cut-off and thus of $T_K$ has a
transparent physical reason.  Indeed, the conventional Kondo
Hamiltonian assumes infinitely quick spin flip.  Therefore, the
mapping of the DWP problem to the Kondo problem can be valid {\em
  only for energies below} $\hbar /t_{\text{tun}}$ - inverse
time, which the tunneling process takes.  Otherwise one could
imagine that the Kondo effect takes place for a heavy particle in an
arbitrary potential, not necessarily a double-well one.  The
reason for the absence of the Kondo effect for, e.g., harmonic
oscillator is that the spacing between the first two levels,
$\Delta$, is of the same order as $\varepsilon_3 \sim \hbar/
t_{\text{tun}}$.  Only the energy interval $\Delta \ll
\varepsilon \ll \varepsilon_3 $ is available for the formation of
the logarithmic singularity!  The fact that $\varepsilon_F
t_{\text{tun}}\gg \hbar$ follows directly from the large mass of
the tunneling particle.  Indeed, the magnitude of DWP should be
of the usual atomic order $\sim 1$ V.  At the same time, the
atomic mass exceeds the mass of an electron by about four orders
of magnitude.  Accordingly the tunneling time of an atomic
particle, $t_{\text{tun}}$ should be about two orders of
magnitude longer than the electron tunneling time scale $\sim
\hbar /\varepsilon_F$.  {\em It is slow tunneling of heavy
  particles (rather than some electron-hole symmetry) that causes
  the suppression of the Kondo temperature.}

The authors of Ref.~\onlinecite{zzcom} try to revitalize 2CK
model by { including} finite electron-hole asymmetry.  Up to
our understanding, they still do not fully appreciate the fact
that the proper cut-off energy { cannot exceed} $\varepsilon_3
\ll \varepsilon_F$

Below we {use an approach, which differs from
  Ref.~\onlinecite{ourcom} and} show that {both} electron and
hole contributions taken separately { diverge
  logarithmically}, the ultraviolet cutoff { being}
$\varepsilon_3$ and, therefore, the prefactor { in the
  expression for $T_K$ has nothing to do with electron-hole
  asymmetry}.

The second order corrections to the electron-assisted ($\delta
A_{e}$) and hole-assisted ($\delta A_{h}$) tunneling amplitudes
can be expressed through Matsubara Green functions of electrons and
movable atoms, $G_{e}(\tau)$ and  $G_{\text{at}}(\tau;x_1,x_2)$, respectively:
\begin{equation}
\delta A_{e,h} \simeq  \int_0^{1/  T}\! d \tau\,
G_{\text{at}}(\tau; -a,a) G_{e}(\pm \tau)\, . 
\label{1}
\end{equation}  
Here we omitted numerical prefactors, put $\hbar=1$ and 
used the fact that $G_{e}(\tau)= G_{h}(-\tau)$. 
We also ignored the spatial coordinates and spin indices in 
$G_{e}(\tau)$.  
The Green function of the movable atom,
$G_{\text{at}}(\tau;x_1,x_2)$,  is  determined by the equation 
\begin{equation}
\left[\frac{ \partial_{x}^2}{2M}-{\partial}_\tau  - U(x)\right]
G_{\text{at}}(\tau;x,x') = \delta(\tau)\, \delta(x-x')\, .
\label{2}
\end{equation}
We assumed for simplicity that DWP, $U(x)$, is symmetric, i.e.,
$U(-x)=U(x)$, and has the two minima at $x=\pm a$.  At $\tau < 0
$ the Green function, $G_{at}(\tau)$, vanishes at $\tau < 0$,
because, according to the model, there is only one movable
defect.  The asymptotic behavior of the electron Green function
at $\epsilon_F\tau \gg 1$ is (omitting non-essential for present
discussion constant factors)
\begin{equation}
  G_{\text{el}}(\tau) = \nu(\mu) \frac{\pi T}{\sin \pi T \tau} + \frac{d
    \nu}{d \mu}\left(\frac{\pi T}{\sin \pi T \tau}\right)^2 {\rm
   sign} \, \tau + \dots
\label{3}
\end{equation}
where $\nu(\mu)$ is the density of states at the Fermi level.  
The second term manifests the electron-hole asymmetry, whatever 
it means.  Notice, that when $\tau$ increases, the 
second term decays {\em faster} than the first one. 
Therefore, this asymmetry becomes irrelevant at large enough $\tau$.
The $\tau$ -dependence of the DWP Green function, $G_{\text{at}}$, 
(\ref{2}) can be found in semiclassical approximation, which is valid for 
excited states as well as for the exponentially weak tunneling: 
\begin{equation} 
G_{\text{at}}(\tau) = B(\tau) \exp\left[-
  S_{\text{cl}}(x_1,x_2,\tau)\right]\, , 
\label{4} 
\end{equation} 
where $S_{\text{cl}}$ is the classical action of the particle 
in the inverted DWP:
\begin{eqnarray}
&&S_{\text{cl}}(x_1,x_2,\tau)= \int_0^\tau d\tau_1 
\left[
\frac{M}{2}\left(\frac{\partial x}{\partial \tau_1}\right)^2 + U(x)
\right];\\
\label{5}
&&{\delta S}/{\delta [x(\tau_1)]}=0, \quad x(\tau_1=\tau) =x_2, \quad
x(\tau= 0 ) =x_1. \nonumber
\end{eqnarray}
One can evaluate the prefactor $B(\tau)$ in Eq.~(\ref{4}).  
However, we do not need it here.  
We substitute $x_{1,2}=\pm a$ and rewrite Eq.(\ref{5}) as 
\begin{equation}
S_{\text{cl}} = -\tau E(\tau) + \int_{-a}^{a} dx
\sqrt{2M[E+U(x)]}\,  .
\label{6}
\end{equation}
Here $E(\tau)$-function is implicitly given by the equation
\begin{equation}
\tau= \int_{-a}^{a}dx \, \sqrt{\frac{M}{2[E+U(x)]}}\, .
\label{7}
\end{equation}
It follows from Eq.~(\ref{7}), that there are two tunneling
regimes, with different dependences $E(\tau)$.  The
crossover takes place at the time scale
\begin{equation}
t_{\text{tun}} \equiv \int_{0}^{a} \!dx
\left[
\sqrt{\frac{2M}{U(x)- U(a)} }
- \frac{1}{a-x} 
\sqrt{\frac{4M}{U^{\prime\prime}(a)}}
\right].
\label{8}
\end{equation}
The quantity
$t_{\text{tun}}\sim\hbar/\varepsilon_{3}$ has a meaning of a
typical time the particle spends under the barrier during the
tunneling process.  Using Eqs.~(\ref{6}) and (\ref{7}) one can
obtain the following asymptotics of $S_{cl}$ at small $\tau$
\begin{equation}
S_{\text{cl}} = \frac{2 M a^2}{\tau} + \tau
\int_{-a}^{a}\frac{dx}{2a}\, 
U(x)  +{\cal O}(\tau^2) \, .
\label{9}
\end{equation}
Equation (9) holds
for $\tau \ll t_{\text{tun}}$. In the opposite case, $\tau \gg
t_{\text{tun}}$, 
\begin{eqnarray}
S_{\text{cl}}& =& - {\tau} U(a) + \int_{-a}^{a}dx
\sqrt{{2M[U(x)-U(a)]}}
\nonumber \\ &&
+ 2 \sqrt{ma^4U^{\prime\prime}(a)}\,
\exp\left[\sqrt{\frac{U^{\prime\prime(a)}}{M}}\left(t_{\text{tun}}-\tau 
\right)\right]\, .
\label{10}
\end{eqnarray}
The first term in Eq.~(\ref{10}) is nothing but the leading in
$\hbar$ term in the energy of the ground state which cancels out
for the on-shell processes.  The second term is the action
corresponding to the tunneling from one minimum to the other.
The classical trajectory for this process corresponds to the kink
with the characteristic time scale $t_{\text{tun}}$. The last
term in the action characterizes the corrections due to the
finite size of the kink (we write it here only for the references
purpose).  Substituting Eqs.~(\ref{3}) and Eq.~(\ref{4}) into
Eq.~(\ref{1}), and using the condition $\epsilon_F
t_{\text{tun}} \gg 1$,  and asymptotic expansions (\ref{9}) and
(\ref{10}) we obtain
\begin{eqnarray}
  &&\delta A_{e,h} =
  \exp\left[-\int_{-a}^{a}\! \! dx\sqrt{{2M[U(x)-U(a)]}} \right] \nonumber
 \\ &&\times
\left\{ \nu(\mu) \left[ \ln \left(\frac{1}{Tt_{tun}}\right) +
      {\cal O}(1)\right] + \frac{1}{t_{\text{tun}}}
\frac{d\nu(\mu)}{d\mu}{\cal O}(1) \right\} .
\label{result}
\end{eqnarray}
 We are not discussing here if the two contributions, $\delta
A_{e}$ and $\delta A_{h}$, add or cancel each other (partially or
completely).  What we would like to emphasize is that for {\em
  each} of the two contributions the {\em cut-off of the
  logarithmic dependence of each contribution is determined only
  by the tunneling time}.  This time exceeds inverse Fermi energy
of the electrons by orders of magnitude, since the atomic mass
$M$ is much bigger than the electron mass.

In fact, this asymmetry (i.e. finite $\partial \nu /\partial
\mu)$ is relevant only at short times $\tau \ll
t_{{\text{tun}}}$, where the integral over $\tau$ in
Eq.~(\ref{1}) is not logarithmic.  We conclude that the origin of
ultraviolet limit of the logarithm has nothing to do with the
electron hole asymmetry.  The smallness of the prefactor in the
expression for the Kondo temperature expresses the simple fact
that for the logarithmic divergence to appear in any physical
property a heavy particle {\em must tunnel} and its tunneling
time is finite, just as it was explained in
Ref.~\onlinecite{ourcom}.  At times smaller than the
characteristic time of tunneling, the heavy particle does not
feel the two-well structure of the potential yet, and does not
acquire any logarithmically divergent correction.  Here we
demonstrated this fact by an explicit calculation alternative to
Ref.  ~\onlinecite{ourcom}. The physical model used in
Ref.~\onlinecite{zzcom} does not differ from and is the same as
the one that was discussed above, of Ref.~\onlinecite{zzcom}.
For these reasons we believe that the calculation is erroneous.

The final remark of Ref.~\onlinecite{zzcom} that the degeneracy
of the low-lying states may be ``guaranteed by some local
symmetry (rotational or possibly time reversal)'' is also, in our
opinion, irrelevant for metallic systems.  Time reversal symmetry
is known to guard the double degeneracy only for systems, which
Hamiltonian contains a half-integer spin (Kramers degeneracy).
Such a Hamiltonian leads to the regular Kondo problem and has
nothing to do with the original two-level system.  As to
rotational symmetries, they cause no degeneracies for the ground
state, because the ground state of a single atom always belongs to the
{\em one-dimensional} representation $A_1$ or $A$ of the point symmetry
group.

To summarize, we disagree with the statement of
Ref.~\onlinecite{zzcom}, that the suppression 
of the Kondo temperature is a consequence of the electron hole
symmetry.
We insist that the Kondo temperature
for a particle in DWP is always parametrically less than the
splitting between the two lowest state.  Accordingly, {\em the
  strong coupling 2CK regime never takes place in metals}.

\widetext

\end{document}